# On analogy between transition to turbulence in fluids and plasticity in solids

S. Ananiev

*Institute of Lightweight Structures and Conceptual Design, University of Stuttgart, Pfaffenwaldring 7, D-70569 Stuttgart, Germany, sergey.ananiev@po.uni-stuttgart.de*

**Abstract -** The subject of this work is the instability mechanism of simple shear flows, like Hagen-Poiseuille pipe flow, which is a long-standing problem in fluid mechanics [1,2]. A possible analogy with phenomenological theory of ideal plasticity in solids is explored. It allows an extension of the Navier-Stokes equations making the simple shear flows unstable.

Following von Mises [4], the existence of maximal allowed shear stress or "yield stress" in fluid is assumed. After the actual stresses have reached this value, a new physical mechanism will be activated attempting to reduce them. This mechanism results in a pressure-like force, which has to be added to the Navier-Stokes equations. The yield stress itself is a material constant, which does not depend on the Reynolds number of particular flow. It will be shown how to estimate its value from experimental data.

To demonstrate how the character of flow changes if the additional force is taken into account, an unsteady flow in a 2D nozzle is presented. The momentum source was introduced in the Navier-Stokes equations through the user-defined function's interface offered by Fluent [6]. The initial data and the results of simulation are summarized in appendix.

## 1. Analogy with plasticity theory

The presence of pressure gradient in transverse to the flow direction is known to be a reason for instability in some types of shear flows, like Taylor-Couette flow [1]. In contrast to that, the reason for instability of shear flows without transverse pressure gradient is still not fully understood [2]. This work attempts to clarify this mechanism by extending the Navier-Stokes equations "forcing" their instability in the case of simple shear flows. An assumption that mechanical behavior of Newton fluids includes also a plasticity-like mechanism resulting in an additional transverse pressure, allows such an extension.

The study of the instability mechanism of shear flows is motivated not only by its great industrial significance, but also by its key role in the turbulence phenomenon [3].

### 1.1. Yield stress

Following von Mises [4] we assume that, similar to metals, shear stresses in Newton fluid can not exceed some value – the yield stress ($\sigma_{xy} \leq \tau_0$), which is a material constant. It seems also physical to assume that this constant has a similar nature as molecular viscosity of fluid ($\mu$), i.e. highly viscous materials possessing a large yield stress as well. In other words, in materials like honey the turbulence is not expected.

To generalize this idea to 2D or 3D case, it is important to guarantee the frame invariance of the plastic yielding criterion. Obviously, for isotropic materials the onset of plastic deformations must be independent of rotation of coordinate system. It is also known that during plastic yielding, metals behave as an incompressible material and therefore the yield criterion must be independent of volumetric pressure. Due to these reasons, von Mises has proposed the distortion energy as a suitable expression for the equivalent stress ($\sigma^{\text{Mises}}$):

$$3\text{D}: \quad \sqrt{\frac{1}{6}\left(\left(\sigma_{xx}-\sigma_{yy}\right)^2+\left(\sigma_{yy}-\sigma_{zz}\right)^2+\left(\sigma_{zz}-\sigma_{xx}\right)^2\right)+\sigma_{xy}^2+\sigma_{yz}^2+\sigma_{zx}^2} \le \tau_0$$

$$2\text{D}: \quad \sqrt{\frac{1}{4}\left(\sigma_{xx}-\sigma_{yy}\right)^2+\sigma_{xy}^2} \le \tau_0 \tag{1}$$

It seems physical to assume, that the transition to turbulence does not depend on volumetric pressure as well. Therefore, the von Mises equivalent stress is adopted in this work.

## 1.2. Additional vorticity

If taken alone, the hypothesis concerning the existence of maximal allowed shear stress is not very helpful. It is necessary to make an assumption about the physical mechanism, which allows reduction of Newton shear stresses to fulfil the von Mises criterion (Eq. 1). Obviously, in the case of the rigid body rotation the shear stresses are equal to zero. This observation suggests an introduction in the model of additional vorticity ($\vec{\omega}^A$), which corresponds to the partial rigid body rotation appearing *spontaneously* in the areas of shear flow where the von Mises criterion is violated. Its counterpart in plasticity is the plastic strain tensor.

For isotropic solids, the rate of plastic strain tensor is coaxial to the elastic strain tensor [4]. It seems physical to assume that in the case of isotropic fluid the vector of additional vorticity is coaxial with the rotor of the instant velocity field:

$$\vec{\omega}^A = \lambda \frac{1}{2}\text{rot}\,\vec{u}, \quad 0 \le \lambda < 1 \tag{2}$$

The scalar $\lambda$ defines the amount of rigid body rotation, which is necessary to reduce the actual Newton stresses to the yield limit $\tau_0$. Its counterpart in plasticity is the Lagrange multiplier. In some cases, where the expression for equivalent stress is simple enough, like von Mises one, its value can be determined analytically from Equation 1. In computational plasticity, this procedure is known as "radial return method" [4]. It projects the elastic stress tensor (in our case Newton stresses) onto the yield surface.

In order to derive an analytical expression for parameter $\lambda$, the gradient of the instant velocity field with additional vorticity must be rewritten in the matrix form. Following our coaxiality assumption (Eq. 2) it reads:

$$\frac{\partial \vec{u}^*}{\partial \vec{x}} = \frac{\partial \vec{u}}{\partial \vec{x}} - \lambda \left[\frac{\partial \vec{u}}{\partial \vec{x}}\right]^{\text{T}}, \quad 0 \le \lambda < 1$$

$$\frac{\partial \vec{u}}{\partial \vec{x}} = \begin{bmatrix} a & b & c \\ d & e & f \\ g & h & i \end{bmatrix} \;\rightarrow\; \frac{\partial \vec{u}^*}{\partial \vec{x}} = \begin{bmatrix} (1-\lambda)a & b-\lambda d & c-\lambda g \\ d-\lambda b & (1-\lambda)e & f-\lambda h \\ g-\lambda c & h-\lambda f & (1-\lambda)i \end{bmatrix} \tag{3}$$

$$\sigma^* = \mu \left[\frac{\partial \vec{u}^*}{\partial \vec{x}}\right]^{\text{Sym}} = (1-\lambda)\sigma^{\text{Newton}}$$

To obtain a new value of stress tensor only the symmetric part of the new velocity gradient has to be taken. It is easy to verify that projected stresses are equal to the Newton stresses multiplied by $(1-\lambda)$. Substitution of these new stresses into the von Mises yield criterion (Eq. 1) leads to the following expression for $\lambda$:

$$\lambda = \begin{cases} \left(1 - \dfrac{\tau_0}{\sigma^{\text{Mises}}}\right) & , \text{if}\left(\sigma^{\text{Mises}} > \tau_0\right) \\[2mm] 0 & , \text{if}\left(\sigma^{\text{Mises}} \le \tau_0\right) \end{cases} \tag{4}$$

### 1.3. Additional force

Now we have to decide how to incorporate the additional vorticity in the momentum conservation equations. Our approach shall not be confused with the theory of micropolar continuum [5], where vorticity is treated as an independent variable. We have assumed the fluid particles do not have any rotary inertia and therefore the conservation of linear momentum only must be considered. This has two practical advantages:

- the stress tensor remains symmetric;
- there is no need in separate boundary conditions for vorticity.

The second Newton's law suggests that the spontaneous generation of additional vorticity has to be interpreted at the continuum level as the result of acting of some centrifugal force:

$$\vec{f}^A = -\rho\,\vec{u} \times \vec{\omega}^A \tag{5}$$

Simple calculations show that the force must indeed depend on the velocity (violating the Galilean invariance!). Due to the spontaneous character of the additional vorticity, the change of velocity component $u_y$ by $du_y$ must be considered as *given*. On the other hand, this change takes place during the time while the particle is being located in the control volume $dv$. This time interval T is $dx/u_x$. According to the second Newton's law, we obtain:

$$du_y = -\int_0^T \frac{f_y^A}{\rho}\,dt \quad \rightarrow \quad du_y = -\frac{f_y^A}{\rho}\frac{dx}{u_x} \quad \rightarrow \quad f_y^A = -\rho u_x \frac{du_y}{dx} \tag{6}$$

Due to its centrifugal character, the additional force has two important properties:

- it does not change kinetic energy of the flow ($\vec{u} \cdot \vec{f}^A = 0$),
- it is identically zero for fluids and flows with $\mu = 0$, $\operatorname{rot}\vec{u} = \mathbf{0}$ or $\vec{u} \times \operatorname{rot}\vec{u} = \mathbf{0}$.

## 2. Estimation of the yield stress $\tau_0$

An accurate determination of the yield stress of some fluid requires a careful evaluation of the experimental data. This is the subject of future work. It is possible however to obtain a rough estimation of this material constant. To this end, the original Reynolds experiment [1,3] has to be interpreted anew taking into account the proposed plasticity-like instability mechanism. It predicts transverse pressure gradient in the wall regions, where the von Mises criterion is violated (Fig. 1).

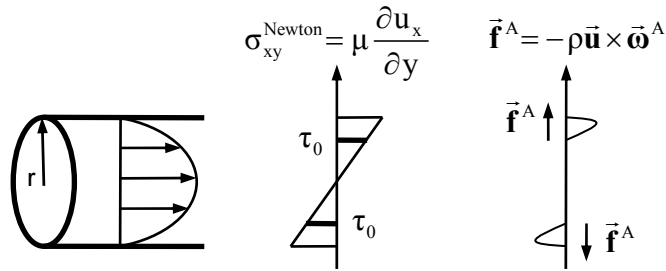

Figure 1. Instability mechanism of Hagen-Poiseuille pipe flow

It is important to understand that if the von Mises criterion is nowhere violated, then the shear flow is stable. This is in a good qualitative agreement with the observations of Reynolds, who found out that if velocity in pipe flow does not exceed some critical value then *all* perturbations of the parabolic flow, large as well as small decay eventually. This allows an estimation of the yield stress of water. The minimal critical Reynolds number (Re = $u^{mean} \cdot d/\nu$) is known to be 2000. The diameter of the one of the Reynolds pipes was approximately 0.02 m. The density of water is 1000 kg/m$^3$ and its molecular viscosity is 0.001 N·sec/m$^2$. From where it follows, that critical mean velocity is 0.1 m/sec. The velocity gradient at the wall is 30 sec$^{-1}$. The corresponding shear stress is **3e-2 N/m$^2$**. This value is assumed to be the yield stress of water.

## 3. Example

To demonstrate how the character of flow changes if the additional force is taken into account, an unsteady flow in a 2D nozzle is presented. An interesting result of this simulation is the flow separation *before* the nozzle, which is not predicted by laminar calculation.

The momentum source was introduced in the Navier-Stokes equations through the user-defined function's interface offered by Fluent [6]. The initial data and the results of the simulation are summarized in appendix. The yield stress of the air was estimated according to our assumption concerning the similarity in physical nature of the molecular viscosity and the yield stress (Sec. 1.1).

$$\frac{\mu^{Water}}{\mu^{Air}} = \frac{\tau_0^{Water}}{\tau_0^{Air}} \quad \rightarrow \quad \tau_0^{Air} \approx 5e\text{-}4 \, N/m^2 \tag{7}$$

## 4. Closure

Detailed study of the physical nature of this plasticity-like mechanism of instability in shear flows is the subject of future work. For the time being, it can be understood as a simple numerical trick, which "forces" the instability of linearly stable flows, like Hagen-Poiseuille pipe flow. The proposed source term can be easily implemented in existing CFD-Software allowing its numerical verification. This makes it potentially useful for practical applications.

## Acknowledgements

Author would like to thank Petr Nikrityuk for motivating discussions concerning turbulence phenomenon and CFD.

**Appendix**

# Unsteady flow through a 2D nozzle ( Re ~ 10$^6$ )

<u>Geometry & grid</u>                    <u>FLUENT settings</u>

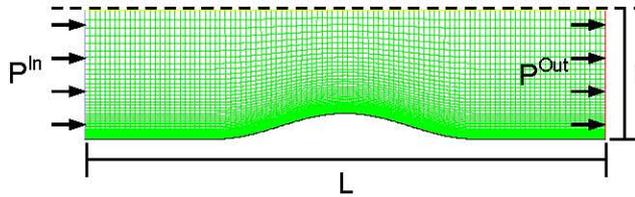

H = 0.1m, L = 0.4m, Nozzle = 10% of area
P$^{In}$ = 0.9 atm, P$^{Out}$ = 0.8 atm
Solver: laminar, coupled, 2d-order implicit
$\rho$ = 1.2 kg/m$^3$, $\mu$ = 1.8e-5 kg/m·sec
u$^{mean}$ ~ 120 m/sec
$\tau_0$ = 5e-4 N/m$^2$

**Dynamics of the von Mises equivalent stress (time interval: 0.01 − 0.0225 sec)**

<u>unsteady Navier-Stokes with additional force</u>          <u>unsteady Navier-Stokes</u>

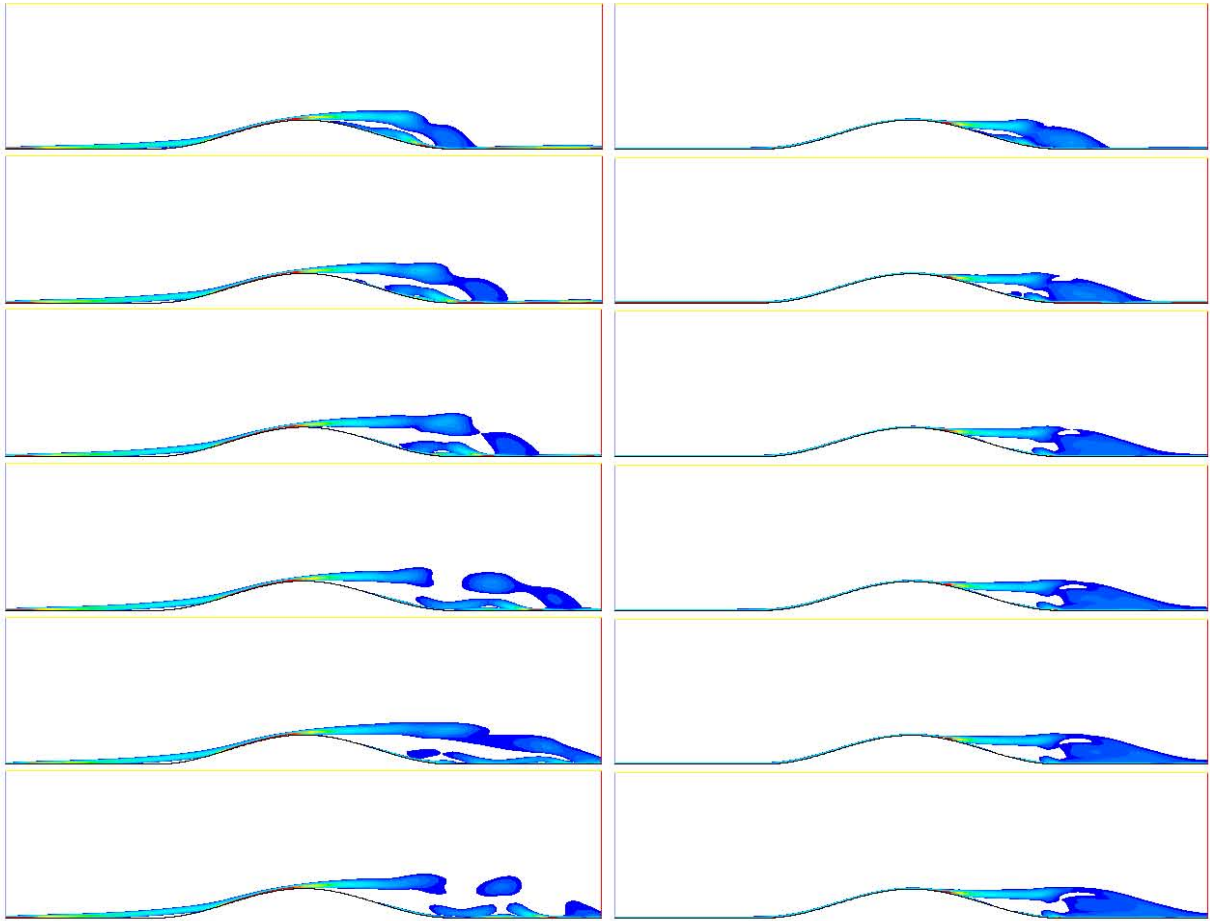

**Velocities at the time increment = 0.0225 sec**

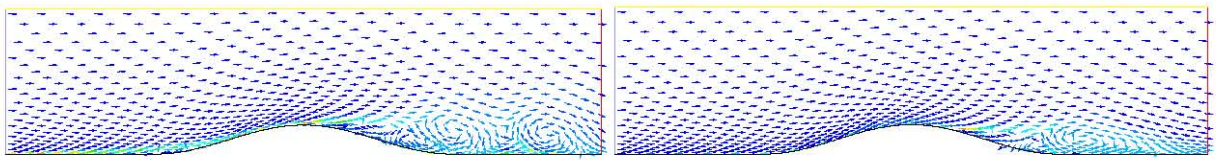